\def\KITAInfPrivFinal{1}
\documentclass{article}

\ifdefined\KITAInfPrivFinal
  \usepackage[dblblindworkshop, final]{neurips_infpriv_2026}
  \usepackage{infpriv_2026_compat}
\else
  \ifdefined\KITAAnonymousWorkshop
    \usepackage[dblblindworkshop]{neurips_2026}
  \else
    \usepackage[preprint]{neurips_2026}
  \fi
  \workshoptitle{NeurIPS 2026 Workshop}
\fi
\ifdefined\KITAAnonymousWorkshop
  \makeatletter
  \renewcommand{\@noticestring}{Submitted to a NeurIPS 2026 Workshop. Do not distribute.}
  \makeatother
\fi

\usepackage{amsmath,amssymb,amsthm}
\usepackage{graphicx}
\usepackage{booktabs}
\usepackage{array}
\usepackage{float}
\usepackage{placeins}
\usepackage{xcolor}
\usepackage{enumitem}
\usepackage{tikz}
\usetikzlibrary{arrows.meta,positioning,fit,backgrounds,calc}
\usepackage{url}
\usepackage[bookmarks=false,colorlinks=true,linkcolor=black,citecolor=black,urlcolor=blue!55!black]{hyperref}

\newcommand{\sk}{\mathsf{sk}}
\newcommand{\pk}{\mathsf{pk}}
\newcommand{\Zq}{\mathbb{Z}_\ell}

\newtheorem{theorem}{Theorem}
\newtheorem{proposition}[theorem]{Proposition}
\newcolumntype{L}[1]{>{\raggedright\arraybackslash}p{#1}}
\definecolor{boxblue}{RGB}{31,92,153}
\definecolor{boxbg}{RGB}{244,247,251}

\title{From Review to Authorization: Key-Isolated Threshold Signing for LLM Agents}

\author{
  Yu Zheng\thanks{Equal contribution.} \\
  University of California, Berkeley \\
  \texttt{yu.zheng@berkeley.edu}
  \And
  Qizhi Zhang\footnotemark[1] \\
  Independent Researcher \\
  \texttt{zqz.math@gmail.com}
}

\begin{document}
\raggedbottom
\maketitle

\begin{abstract}
Autonomous LLM agents can turn semantic judgments into consequential actions
such as payments and permission changes. If the same process makes these
judgments and controls a reusable signing credential, a mistaken judgment
can reach execution without further approval. To make LLM review
govern execution, we present \textsf{KITA}, an authorization architecture
that keeps the user's personal signing key and every threshold signing-key
share outside all LLM processes. Under threshold-signature unforgeability
and our system assumptions, an adversary controlling the proposer and fewer
than $t$ reviewer--signer domains cannot produce a valid authorization for
a new action without signing contributions from $t$ distinct domains. This execution-bound authorization
integrity ensures that each accepted authorization includes a share from an
uncompromised domain, bound to the canonical action and released only after
authenticated reviewer approval.
We implement an in-process prototype of \textsf{KITA} with a structured-output
LLM adapter and threshold BLS. Six system tests validate quorum gating and
message binding, seven controlled baseline scenarios compare authorization
outcomes, and cryptographic microbenchmarks measure online signing cost and
its scaling behavior.

\end{abstract}

\section{Introduction}
Suppose a user asks a personal agent (e.g., Meta Muse~\citep{metamuse2026}, OpenAI Dots~\citep{openaidots2026}) to buy a new laptop for at most
\$1,500 from shopping platforms (e.g., Amazon). The agent reads product pages, compares offers, selects a merchant,
and initiates a payment. These steps mix two different kinds of computation.
Comparing offers is a judgment: the agent interprets information and may be
wrong. Payment is an external action: it changes the user's state and may be
costly or impossible to reverse. This distinction motivates separating
model-based reasoning from execution authority.

The agent may misread a product's condition, rely on misleading evidence, or
encounter indirect prompt injection that changes its plan or tool
calls~\citep{zhan2024injecagent,debenedetti2024agentdojo}. If the same process
also controls a reusable signing credential (e.g., user's personal key), any of these failures can reach
the authorization path. The security question extends beyond the quality of
the decision to whether one LLM-facing process can execute that decision on
its own. Additional reviewers help enforce the user's requirements only if
their approvals govern access to execution authority.

Agent payment protocols already provide part of this boundary. AP2, for
example, binds a concrete purchase mandate to a user-authorized open mandate
and places mandate validation in deterministic code~\citep{ap2}. These
mechanisms establish the scope of a delegation. Our work complements them
by asking how semantic approval of a concrete action can become enforceable
authority within that scope.

\begin{samepage}
Three research lines provide the ingredients for this composition.
Multi-agent debate, voting, and model panels review semantic
decisions~\citep{du2024debate,verga2024juries,choi2025debate}. Policy and
delegation systems constrain privileges and enforce machine-checkable
limits~\citep{debenedetti2025camel,shi2025progent,south2025delegation,muruaga2026bounded}.
Threshold and collective signing make a quorum cryptographically necessary
for a message~\citep{borselius2002,syta2016cosi}. Bringing these mechanisms
together requires a link between the reviewers' decisions, the signing
credential, and the action that the executor actually performs. If the
proposer retains a complete signing key, it can ignore a committee's
rejection; a central signer can enforce the review policy, but keeps the
complete credential in one failure domain.
\end{samepage}

We realize this composition with \textsf{KITA}, a key-isolated threshold
authorization architecture. A proposer with no signing-key share can exchange
feedback with reviewers while selecting a candidate. It then fixes a canonical
action and its supporting evidence for authorization. Reviewer domains assess
the same object against authenticated user requirements. Within each domain,
an isolated signer retains a share of a fresh delegated key and releases an
action-bound signature share upon authenticated approval. The user's
personal key remains outside the agent system. Contributions from $t$
distinct domains form one signature for a compatible executor. Changing an
amount, recipient, permission, or other signed field requires a new review
and quorum.

\textsf{KITA} separates the guarantee that the required parties approved an action
from the question of whether their judgments were correct. Under the stated
system and cryptographic assumptions, a compromised proposer and fewer than
$t$ corrupt reviewer domains cannot recover the delegated key or bypass the
quorum, except with negligible probability. With conditionally independent
reviewer errors and fixed committee parameters, \textsf{KITA} can reduce the per-attempt
probability of unsafe authorization to $O(pq^{t-f})+\epsilon_{\rm crypt}$,
where $p$ is the probability of an unsafe proposal and $q$ bounds each
uncorrupted reviewer's false approvals. We also
analyze reviewer dependence, withholding, and repeated attempts. Our contributions are:
\begin{itemize}[leftmargin=1.5em,itemsep=3pt,topsep=3pt]
  \item A key-isolated authorization architecture that binds review, signing,
  and execution to one canonical action while keeping personal keys and
  threshold signing-key shares outside LLM runtimes
  (Section~\ref{sec:system}).
  \item A composition analysis of key protection and quorum integrity,
  together with conditional bounds on unsafe authorization and timely
  completion under fallible review (Section~\ref{sec:security}).
  \item An in-process prototype connecting structured reviewer outputs,
  authenticated approval gateways, threshold BLS, and an executor.
  Controlled fault tests validate authorization checks, and sensitivity
  simulations examine the trade-offs in review quality
  (Section~\ref{sec:eval}).
\end{itemize}

\section{\textsf{KITA}: From Review to Authorization}
\label{sec:system}
\textsf{KITA} has four roles: a human user $U$, a personal-agent proposer $P$, reviewer
domains $R_1,\ldots,R_n$, and a verifier/executor $V$. In a shopping workflow,
$P$ searches for products, cloud review services assess the proposed purchase,
and $V$ is the merchant or payment endpoint that executes the order. Each
reviewer domain contains an LLM and an isolated signer $S_i$; the domain is
the unit of corruption in our security model. Figure~\ref{fig:arch} shows
how these roles interact.

Before the agent begins, $U$ authenticates the review policy $\pi$ to the
reviewers and $V$. Binding requirements, such as a budget or a requirement
for a new product, belong in this policy. Preferences sent only to $P$ can
guide search, but reviewers need an authenticated reference to enforce them.
The user retains a personal key $\sk_u$ and provisions shares of a fresh
delegated key $\sk$, with threshold $2\le t\le n$. The committee's public key
$\pk$ and key epoch are also authenticated. Appendix~\ref{app:lifecycle}
describes provisioning with erasure of the complete delegated key and the
alternative of distributed key generation.

\begin{figure}[ht]
\centering
\begin{tikzpicture}[
  font=\small,
  box/.style={draw=boxblue,rounded corners=2pt,align=center,
    inner sep=4pt,minimum height=8mm},
  worker/.style={box,fill=boxbg,minimum width=22mm,minimum height=6mm},
  signer/.style={box,fill=white,minimum width=22mm,minimum height=6mm},
  flow/.style={->,thick,boxblue},
  setup/.style={->,dashed,black!65},
  lbl/.style={font=\footnotesize,fill=white,inner sep=2pt},
  >={Stealth[length=2mm]}]
\node[box,fill=black!4] (user) at (0.3,0) {User $U$};
\node[worker] (prop) at (3.3,0) {Proposer $P$\\{\footnotesize search / combine}};
\node[box,fill=black!4] (ver) at (11.2,0) {Verifier $V$\\{\footnotesize verify / execute}};
\foreach \i/\x in {1/4,2/7,n/10} {
  \node[worker] (r\i) at (\x,2.8) {LLM review};
  \node[signer] (s\i) at (\x,1.8) {Isolated $S_{\i}$: $\sk_{\i}$};
  \draw[flow] (r\i) -- (s\i);
  \begin{scope}[on background layer]
    \node[draw=boxblue,dashed,rounded corners=3pt,
      fit=(r\i)(s\i),inner sep=4pt,
      label={[font=\small]above:domain $R_{\i}$}] (d\i) {};
  \end{scope}
}
\node at (8.5,2.3) {$\cdots$};
\node[lbl] at (7.65,2.3) {approve};
\node[fit=(d1)(d2)(dn),inner sep=1pt] (committee) {};
\draw[setup] (user) -- node[lbl,above]{request} (prop);
\draw[setup] (user.north) to[out=85,in=180]
  node[lbl,pos=.5,left]{policy / setup} (committee.west);
\draw[flow] (prop.north west) to[out=115,in=245]
  node[lbl,pos=.4,left]{$m,E,\pi$} (committee.south west);
\draw[flow] (committee.south) to[out=-90,in=60]
  node[lbl,pos=.45,right]{$\sigma_i$ on $m$} (prop.north east);
\draw[flow] (prop) -- node[lbl,above]{$m,\sigma$ from $t$ distinct domains} (ver);
\draw[setup] (user.south) -- (0.3,-0.9)
  -- node[lbl,below]{$U$-authenticated policy, public key, epoch} (11.2,-0.9)
  -- (ver.south);
\end{tikzpicture}
\caption{Review and key custody occupy separate components within each
domain. An authenticated approval activates the paired signer; only
action-bound signature shares leave it. The proposer combines a quorum into
one signature, and the verifier executes the signed action. The diagram shows
the authorization path after candidate selection and reviewer feedback. Dashed arrows
denote user inputs and setup; solid arrows denote the authorization path.}
\label{fig:arch}
\end{figure}

\paragraph{From a proposal to a reviewed action.}
Before requesting authorization, $P$ can discuss candidates with the reviewers
over multiple rounds. Reviewers may identify a policy mismatch, request
evidence, or suggest changes; $P$ uses this feedback to refine its search.
For example, a reviewer may flag a \textit{like-new, renewed} laptop and ask
$P$ to find a new unit. These exchanges produce feedback without activating
the signers.

Once $P$ selects a candidate, it freezes a canonical action $m$: an  encoding of the order, evidence digest, policy reference,
execution domain, request and attempt identifiers, nonce, and expiry.
Each reviewer receives this package for a final authorization decision and
checks it against $\pi$. Deterministic checks enforce structured constraints
such as price and payee, while the LLM interprets product condition and
compatibility evidence. A change to the order or supporting evidence returns
the candidate to revision and requires a fresh authorization attempt.
Appendix~\ref{app:review-task} develops the purchase example.
Only a final approval of the exact bytes of $m$ enables signing. A gateway
outside the LLM runtime authenticates that decision to the paired signer, which releases
$\sigma_i=\mathsf{SignShare}(\sk_i,m)$. 
\paragraph{Enforcing the quorum.}
The proposer, or another untrusted combiner, verifies shares from $t$ distinct
domains on the same message and combines them into a signature $\sigma$.
$V$ checks this signature and the execution constraints, then acts on the
signed fields. For a one-purchase request, all attempts share a user-authenticated
request ID. Execution atomically consumes that ID and the nonce, or uses an
equivalent idempotency mechanism. This prevents both replay of one authorization
and execution of several authorized revisions of the same purchase.
The review harness durably caches each domain's first decision for a message.
A retry returns the cached result; a revision or a fresh review requires a
new signed attempt ID and nonce. 
Appendix~\ref{app:object} gives
the canonical object and state machine, and Appendix~\ref{app:boundary}
specifies the service boundaries.

\paragraph{Why threshold signing?}
A central signer can enforce the same review policy, but compromise of that
service exposes a complete key. Threshold signing removes this complete-key
custodian from the online path. It also gives a compatible executor one
public key and one signature to verify, regardless of committee size.
An executor that can verify multiple authenticated approvals can instead
enforce the quorum directly, with equivalent protection under corresponding
domain assumptions. Preserving this single-signature interface comes with the cost of distributed
key setup and share management. Appendix~\ref{app:baselines} compares the
credential placement and failure boundaries of these designs, and
Appendix~\ref{app:ap2} discusses payment integration and signature compatibility.

\section{Security Model and Main Results}
\label{sec:security}
The analysis follows the separation in the architecture. We first ask whether
an adversary can recover a reusable key or authorize an action without a
quorum. We then ask how often a correctly enforced quorum may approve an
action that violates the user's requirements.

\paragraph{Adversary and Assumptions.}
\textsf{KITA} includes agentic components $P$ and $R_1,\ldots, R_n$. Within one key epoch, an adversary may control $P$, the
evidence supplied to all reviewers, and a fixed set of $f<t$ reviewer domains.
Full corruption reveals a domain's key share and permits arbitrary approvals
or withholding. The remaining domains preserve their harness and signer
boundaries, although their LLMs may make mistakes for any reason such as misinterpreting a product.

\paragraph{Authorization integrity.}
Cryptographic signing connects a review decision to the bytes submitted for execution.
Together with the threshold scheme's security assumptions, they give the
following result. Let $\epsilon_{\rm crypt}(\lambda)$ denote the negligible
aggregate probability of cryptographic failure.
\begin{theorem}[Key protection and quorum integrity]
\label{thm:integrity}
Under the system assumptions above and the threshold scheme's key-recovery and per-message
unforgeability assumptions, any polynomial-time $\mathcal A$ controlling $P$
and $f<t$ domains recovers the complete delegated key only with negligible
probability in security parameter $\lambda$. Except with probability
$\epsilon_{\rm crypt}(\lambda)$, if $V$ accepts a fresh (not previously accepted) action $m$, at least
$t-f$ uncorrupted domains approved those exact bytes and released shares.
\end{theorem}

Theorem~\ref{thm:integrity} holds even when the uncorrupted reviewers make
mistaken approvals: their signers still retain the secret shares and sign
only the approved action. Thus the adversary may exploit review errors to
obtain a particular authorization, while remaining unable to recover the
delegated key or bypass the quorum. The user's personal key is outside the
agent system throughout. Appendix~\ref{app:security} gives the formal model
and proof.

\paragraph{Unsafe authorization under fallible review.}
An action is unsafe when it violates the user's authenticated requirements
in the relevant world state. A final review attempt fixes a candidate $m$,
world state $w$, and history $h$ before reviewer randomness is sampled.
The history includes all earlier search, feedback, and review interactions,
including the multi-round discussion that led $P$ to select this candidate.
The relevant world-state conditions must hold through execution or be
revalidated, as specified in Appendix~\ref{app:security}.

Let $p$ be the probability of an unsafe proposal.
For each uncorrupted domain, let $X_i=1$ denote a false approval and released
share. We assume
$\Pr(X_i=1\mid m,w,h,\text{unsafe})\le q$
uniformly over eligible unsafe candidates, world states, and histories.
An average error rate over a task distribution can conceal inputs on
which all reviewers fail; Appendix~\ref{app:measurement} gives a counterexample.

Outside a cryptographic failure, an unsafe authorization needs at least
$t-f$ false approvals from the $n-f$ uncorrupted domains, even if every corrupt
domain signs. This counting argument yields the next result, where
$T_{N,k}(z)=\Pr[\operatorname{Bin}(N,z)\ge k]$ is the binomial upper tail.
\begin{theorem}[Unsafe authorization]
\label{thm:semantic}
Let $F$ be the event that $V$ accepts an unsafe action in one attempt, and let
$N=n-f$, $k=t-f$. If the $X_i$ are independent conditional on the candidate,
world state, and history, then
\begin{equation}
 \Pr(F)\le p\,T_{n-f,t-f}(q)+\epsilon_{\rm crypt}.
 \label{eq:unsafe-independent}
\end{equation}
With only the same conditional marginal bounds and arbitrary dependence,
\begin{equation}
 \Pr(F)\le p\min\!\left\{1,\frac{(n-f)q}{t-f}\right\}
             +\epsilon_{\rm crypt}.
 \label{eq:unsafe-dependent}
\end{equation}
\end{theorem}

The two bounds show why the joint behavior of reviewers matters. Under
unanimity, conditional independence gives an unsafe-acceptance bound of
$pq^{n-f}+\epsilon_{\rm crypt}$. Fully shared errors can instead yield $pq$:
adding reviewers then provides no additional semantic protection. The gap between these bounds shows that committee gains depend on the reviewers' joint error behavior. Appendix~\ref{app:security} proves both bounds.

\paragraph{Completion and repeated attempts.}
A stricter threshold improves system safety but also makes valid actions harder to complete. For every
valid candidate and history, let $r$ bound each uncorrupted domain's
probability of missing the approval deadline. Under conditional independence, a valid candidate has enough shares
by deadline $D$ with probability at least $T_{n-f,t}(1-r)$ despite corrupt
withholding. If subsequent forwarding, combination, and execution take at
most $\Delta$ and the request remains eligible, the same bound guarantees
completion by $D+\Delta$. Jointly tolerating $f$ corrupt domains for
integrity and worst-case availability therefore requires $f<t\le n-f$.
The nonce checks prevent replay, but a fresh review gives an adversarial proposer another opportunity to
obtain enough mistaken approvals. For a bounded attempt budget, a union bound
sums the per-history unsafe-acceptance bounds. The cached-decision rule in
Section~\ref{sec:system} makes each fresh set of reviews a distinct signed
attempt. Appendices~\ref{app:availability} and~\ref{app:retries} give the
completion and multiple-attempt results.

\section{Validation and Trade-offs}
\label{sec:eval}
Appendix~\ref{app:benchmarks} reports microbenchmarks of the prototype's
signing operations.

\paragraph{Authorization checks.}
In six scripted system tests (Appendix~\ref{app:system-tests}), our
in-process $2$-of-$3$ BLS prototype accepts a quorum-approved action, rejects
insufficient quorum, mismatched digests, post-review substitution, and replay,
and gives reviewers identical, key-free inputs. Across seven
scripted scenarios, \textsf{KITA} and direct approval verification produce the same
outcomes: one exposed reviewer credential remains below quorum. Exposing a central signer's complete key bypasses its approval check. These outcomes reflect the credential placement discussed in Section~\ref{sec:system}: distributing custody
protects against a single credential failure.
A benign purchase request reviewed by three model calls also completes the
path from proposal to execution. Final verification at $V$ is independent of
committee size; the combiner's share checks grow linearly in $t$. Appendix~\ref{app:evaluation}
describes the integration test, fault scenarios, and baseline outcomes.

\begin{figure}[ht]
\centering
\includegraphics[width=0.70\linewidth]{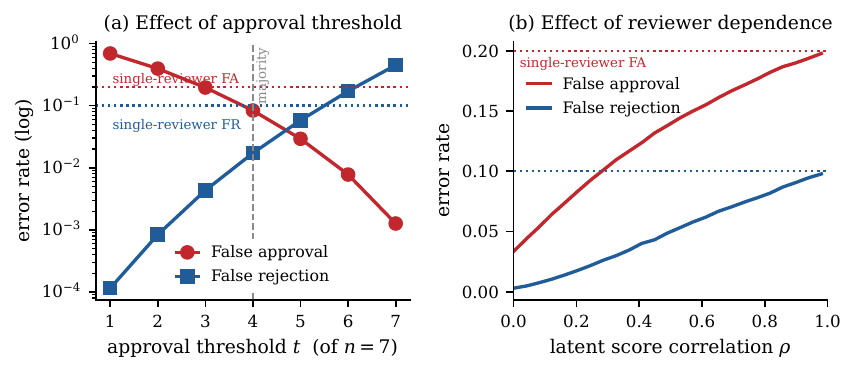}
\caption{\textbf{Quorum size trades safety for completion; shared errors erode
committee gains.} Gaussian sensitivity model with $n=7$, $f=0$, and
marginal rates $\bar q=0.20$ and $\bar r=0.10$: (a) threshold sweep at latent correlation $\rho=0.2$;
(b) dependence sweep at $t=4$. Each estimate uses $400{,}000$ trials.}
\label{fig:committee}
\end{figure}

\paragraph{Threshold and reviewer dependence.}
To study the semantic trade-off, we use a Gaussian sensitivity model. Figure~\ref{fig:committee} varies the
approval threshold and the dependence between reviewers while holding each
reviewer's marginal false-approval and false-rejection rates at
$\bar q=0.20$ and $\bar r=0.10$; these averages are not the uniform bounds
of Section~\ref{sec:security}.

At latent correlation $\rho=0.2$, a $4$-of-$7$ quorum yields false-approval
and false-rejection rates of $0.083$ and $0.017$. Raising the threshold to
unanimity reduces false approval to $0.0013$ but raises false rejection to
$0.45$. Holding $t=4$ instead, false approval rises from $0.033$ under
independence toward $0.20$ as errors become fully shared. In this model,
stricter review and less dependent review have different effects: the former
trades acceptance of unsafe actions for rejection of valid ones, while the
latter can improve both. Appendix~\ref{app:simulation} specifies the model
and reports Monte Carlo uncertainty.

\section{Conclusion}
\label{sec:conclusion}
We explore \textsf{KITA} as a starting point for making collective agent judgment an enforceable authorization boundary. By keeping reusable secret-key material outside LLMs and requiring $t$ approvals on
the exact action, \textsf{KITA} separates semantic review, credential custody, and execution for autonomous agents. Future work can improve the review layer through heterogeneous reviewer ensembles, calibrated confidence, and abstention.

\clearpage
{\footnotesize
\setlength{\bibsep}{1pt}
\bibliographystyle{plainnat}
\bibliography{references}
}

\clearpage
\appendix

\section{Protocol Specification}
\label{app:object}

\subsection{Canonical Authorization Object}

Table~\ref{tab:message} specifies the authorization object serialized as $m$.
Application profiles may extend the schema, but review, signing, and execution
use the same versioned canonicalization algorithm. The evidence manifest $E$
records source identifiers, content digests, retrieval times, and freshness
metadata. Its commitment $H(E)$ binds the reviewed evidence package to $m$.
Material sources used to justify final approval must be included in this
manifest. During final review, re-fetching verifies the committed content
and its freshness. If a reviewer needs changed or additional material evidence,
it requests revision; $P$ updates $E$ and freezes a new $m$ for another attempt.
Earlier discussion may guide the review, but any factual evidence relied on
for authorization must be represented in the frozen package.

\begin{table}[ht]
\centering
\small
\setlength{\tabcolsep}{4pt}
\begin{tabular}{L{0.23\linewidth}L{0.30\linewidth}L{0.37\linewidth}}
\toprule
Field group & Example fields & Why it is signed \\
\midrule
Protocol context & domain separator, schema version & Prevents a signature from being reused in another protocol or parser. \\
Principal and committee & principal ID, committee ID, group-key ID, key epoch & Binds the action to one delegator and the currently authorized share holders. \\
Policy and request & policy ID/version/digest, request ID, attempt budget & Binds user requirements and request-level execution and review limits. \\
Action & type, normalized parameters, execution audience & Defines exactly what the verifier may execute and where it may execute it. \\
Evidence & manifest digest $H(E)$, source types, freshness requirements & Prevents evidence substitution while leaving source validation to reviewers. \\
Freshness and replay & attempt ID, nonce, issue time, expiry & Separates final review attempts and limits authorization lifetime. \\
Linkage & parent proposal digest, external object digest & Binds a revision or an AP2 mandate to the object reviewed by the committee. \\
\bottomrule
\end{tabular}
\caption{Minimum content of the canonical authorization object.}
\label{tab:message}
\end{table}

\paragraph{Canonicalization and parsing.}
The deployment selects one encoding, for example a versioned profile of the JSON Canonicalization
Scheme~\citep{rfc8785}, and specifies field ordering, number representation, string encoding,
and the treatment of absent values. Reviewers and $V$ parse the object, encode it
again, and reject it unless the bytes match the canonical form. Unknown critical
fields, duplicate keys, invalid timestamps, and values outside the application
schema fail closed. The executor consumes the parsed signed fields directly,
binding execution to the verified object through the final state transition.

\paragraph{Request and attempt registration.}
For a one-shot purchase, $U$ authenticates a request ID, policy, and maximum
number $M$ of final authorization attempts to $V$ and the reviewers.
Before final review, $V$ atomically registers a unique binding
$(\mathsf{request\_id},\mathsf{attempt\_id},H(m))$ for the pending request.
Registration of an existing binding is idempotent; a conflicting digest or a
new binding beyond the budget is rejected. Each uncorrupted gateway verifies
this binding through an authenticated channel to $V$ before initiating final
review. This registry holds no signing key. Request state and attempt counts
persist across revisions, policy updates, and key rotations. A new request ID
requires new user authorization.

\subsection{Review and Execution State Machine}
The state machine refines the protocol in Section~\ref{sec:system}, including
the treatment of revisions and retries. Each transition records the proposal
digest and policy/key epoch.
\begin{enumerate}[leftmargin=1.6em,itemsep=2pt,topsep=2pt]
  \item \textbf{Search $\leftrightarrow$ feedback.} $P$ exchanges candidate
  products, questions, and supporting material with reviewers. The harness
  keeps these exchanges in a feedback mode that cannot activate a signer.
  \item \textbf{Candidate $\rightarrow$ final review.} $P$ freezes one canonical
  $m$, registers its attempt with $V$, and sends the same bytes, evidence
  manifest, and policy reference to every $R_i$.
  \item \textbf{Review $\rightarrow$ reject, revise, or approve.} A reviewer
  loads the user-authenticated policy, checks deterministic constraints,
  validates the frozen evidence package, and emits an authenticated outcome
  tagged as a final authorization decision. The gateway checks the registered
  attempt and this decision type before authenticating an approval.
  Only approve$(m)$ causes its isolated signer $S_i$ to produce
  $\sigma_i=\mathsf{SignShare}(\sk_i,m)$; the reviewer's LLM never receives
  $\sk_i$.
  The harness atomically records the first decision for each message digest and
  returns that result on retries, without another model call.
  \item \textbf{Revise $\rightarrow$ new review.} Conditions such as a lower
  amount, narrower table scope, shorter expiry, or a request for a fresh decision
  return the candidate to discussion. The next final review registers a new
  object $m'$ with a fresh attempt ID, nonce, and parent digest under the same
  request ID. No share on $m$ is reused.
  \item \textbf{Approve $\rightarrow$ quorum.} The combiner authenticates the
  claimed reviewer index and verifies each signature share under $\pk_i$. It deduplicates
  indices and combines only shares on the identical digest and epoch.
  \item \textbf{Quorum $\rightarrow$ execute or expire.} $V$ validates the final
  signature, schema, audience, policy version, key epoch, expiry, registered
  attempt, and pending request. Execution atomically consumes the request ID
  and nonce, or uses an equivalent idempotency protocol. Other attempts for
  that request then fail, even if they carry distinct nonces and valid signatures.
\end{enumerate}

Authenticated reject and revise messages preserve the provenance of reviewer
feedback. Audit records can also include rationales in approval envelopes
bound to the digest of $m$. Including a rationale in the signed action itself
requires fixing that field before review and signing.

\subsection{Authorization Alternatives}
\label{app:baselines}
Key isolation alone can be achieved with a single remote signer. The further
choice is where to enforce the review quorum and how much credential custody
to place in one failure domain. Table~\ref{tab:baselines} compares four designs
under a trusted $V$ and authenticated key registration.

\begin{table}[ht]
\centering\small
\setlength{\tabcolsep}{4pt}
\begin{tabular}{L{0.23\linewidth}L{0.29\linewidth}L{0.40\linewidth}}
\toprule
Design & Credential custody & Quorum enforcement and failure boundary \\
\midrule
Agent-held key & Complete key accessible to $P$ & $P$ can sign any action within the delegation. \\
Isolated central signer & Complete key in one remote service & Service checks $t$ approvals; its compromise bypasses that check. \\
Direct approvals at $V$ & Separate reviewer authentication keys & $V$ checks $t$ identities and action bindings; fewer than $t$ credentials cannot bypass the quorum. \\
\textsf{KITA} & One key share per reviewer domain & Signers gate shares; $V$ verifies one group signature. Fewer than $t$ exposed shares cannot bypass the quorum. \\
\bottomrule
\end{tabular}
\caption{Authorization baselines. All quorum designs remain subject to mistaken reviewer approvals.}
\label{tab:baselines}
\end{table}

Public share combination removes the complete-key custodian from \textsf{KITA}'s
online path. Its credential-compromise tolerance matches direct
multi-approval verification under corresponding domain assumptions. The latter
is a natural choice when $V$ supports reviewer identities and quorum checks;
\textsf{KITA} targets a one-key, one-signature execution interface. Both require multiple
review services, while \textsf{KITA} adds distributed key setup and share management.
A central signer has simpler key management but remains a single credential
failure point. Our BLS prototype requires a BLS-capable $V$; preserving an ECDSA
interface requires threshold ECDSA (Appendix~\ref{app:ap2}).

\section{Formal Security Model and Proofs}
\label{app:security}

\subsection{Parties, Setup, and Adversarial View}

There are four role types, with $n$ instances of the reviewer role.
$U$ is a trusted human principal, $P$ is the personal agent and may also combine
shares, $R_i$ is a cloud review domain, and $V$ is the trusted effectful endpoint.
The LLM, orchestration harness, approval gateway, and isolated signer $S_i$
are components of $R_i$. The reviewer evaluates the action; the signer checks
the approval's authentication and digest. An authenticated approve decision
activates the signer, including when that decision is mistaken.

Write $I_\pi(m,w)=1$ when action $m$ satisfies the user's authenticated
requirements in the relevant world state $w$. Reviewers assess the evidence
needed to evaluate this predicate; the executor enforces the signed action
and its machine-checkable constraints.

$U$ authenticates the review policy $\pi$, binding request requirements,
committee identities, threshold $2\le t\le n$, group key $\pk$, and epoch.
Reviewers and $V$ obtain these through authenticated channels or verify a
user-signed delegation using $\pk_u$. Each binding preference belongs in these
authenticated requirements, which define intent through $I_\pi(m,w)$.
Additional preferences supplied only to $P$ guide its search.

Setup installs shares of a fresh delegated key in protected signer stores and
authenticates the policy, epoch, and public keys to all parties. In the dealer
profile, erasure removes the complete delegated key and recoverable setup
state; DKG provides an alternative. Appendix~\ref{app:lifecycle} describes both
procedures. Trusted setup, the user's personal key $\sk_u$, and the user wallet
are outside the adversary's reach.

The corruption set $C$ is fixed for the analyzed epoch, $|C|=f<t$.
$\mathcal A$ controls $P$, corrupt domains, scheduling, and external evidence,
including a common adversarial input to every reviewer.
A conservative domain corruption reveals $\sk_i$ and channel credentials and
allows arbitrary signing, approvals, or withholding. An uncorrupted domain
preserves its process and key boundary while making semantic errors from any
cause. The fixed-set assumption bounds exposed shares over the entire epoch.

$\operatorname{View}_{\mathcal A}$ includes public keys and policy, exposed
shares, proposer state, chosen requests, review feedback, approval records,
signature shares, final signatures, and public execution receipts.
Uncorrupted secret shares and private signing randomness remain hidden;
product evidence and reviewer rationales may be public.
A common administrator or runtime able to expose several shares
must be counted as compromising all of those domains.

\subsubsection{Isolation and domain corruption}
\label{app:boundary}
A deployment can place each review worker in a sandbox with read-only policy
access and allowlisted retrieval tools. A gateway outside that sandbox
validates its structured decision, caches the first result per digest, and
authenticates approvals to the paired signer over a service channel. Channel
credentials and signing shares belong to the gateway and signer services;
the worker and proposer have no direct signing endpoint. Separate service
identities and administrative domains define the corruption units.
Policy and key registration use a user-authenticated administration path.
Table~\ref{tab:failure-map} maps concrete failures to this model. The prototype
implements these interfaces within one process, with HMAC approval authentication.

\begin{table}[ht]
\centering\small
\setlength{\tabcolsep}{4pt}
\begin{tabular}{L{0.40\linewidth}L{0.52\linewidth}}
\toprule
Failure & Treatment in the security model \\
\midrule
Misleading input causes a wrong reviewer decision & False approval $X_i=1$; its signer can release a valid share while retaining key secrecy. \\
Proposer process is controlled & Adversary chooses proposals, evidence, and submissions; worst-case $p=1$. \\
Worker/gateway takeover permits arbitrary approvals & Count the approval domain in $f$; full-domain corruption conservatively grants its key share as well. \\
Signer store or host is compromised & Its share is exposed and the domain counts in $f$. \\
Shared administration exposes several domains & Count every affected domain; separate model calls do not create separate administrative boundaries. \\
Verifier or user setup is compromised & Violates the trusted setup/execution assumptions. \\
\bottomrule
\end{tabular}
\caption{Mapping deployment failures to semantic error, domain corruption, and trusted components.}
\label{tab:failure-map}
\end{table}

\subsection{Security Events and Cryptographic Composition}

The trusted verifier accepts only registered canonical messages bound to the
authorized principal, request and attempt IDs, policy digest/version, audience,
key epoch, expiry, and nonce.
It executes those fields without consulting mutable proposer state.
For a one-shot request, consumption of the request ID and nonce and execution
are atomic or use an equivalent idempotency mechanism. Environmental assumptions needed by $I_\pi$ must remain valid through
execution, or the profile must require revalidation.

For a message $m$, let $Q(m)$ be the set of uncorrupted domains that have
authenticated approve$(m)$ and released a valid share on those exact bytes.
Consider these events:
\begin{itemize}[leftmargin=1.5em,itemsep=2pt,topsep=2pt]
 \item $\mathsf{KeyRecover}$: $\mathcal A(\operatorname{View}_{\mathcal A})$
 outputs a secret key matching the registered delegated public key.
 \item $\mathsf{Bypass}$: $V$ accepts a fresh $m$ with $|Q(m)|<t-f$. We call
 $m$ fresh if $V$ has not previously accepted it; repeated acceptance of an
 accepted $m$ is excluded separately by nonce consumption.
 \item $\mathsf{Unsafe}$: $V$ accepts $m$ with $I_\pi(m,w)=0$.
\end{itemize}
Key recovery and quorum bypass violate the authorization boundary.
Unsafe execution can also result from mistaken reviews with $|Q(m)|\ge t-f$
and correctly functioning cryptography.

Assume a correctly generated threshold scheme with computational key-recovery
resistance and per-message threshold unforgeability under $f<t$ corruptions
and chosen-message access to uncorrupted signers. In the latter game, producing
a valid signature on $m$ is a forgery when fewer than $t-f$ uncorrupted signers
were queried on $m$; access on other messages is allowed.
Also assume unforgeable approval and user-delegation authentication,
collision-resistant commitments, authenticated public shares, and correct
canonicalization and verification. These assumptions support Theorem~\ref{thm:integrity} in
Section~\ref{sec:security}.

\begin{proof}[Proof of Theorem~\ref{thm:integrity}]
In the dealer profile, fewer than $t$ Shamir shares alone are statistically
independent of the shared secret. With public keys and signing transcripts
included, the full-view key-protection claim is computational and follows
from the threshold scheme's key-recovery assumption.
The trusted setup and protected storage ensure that $\mathcal A$ sees no
additional secret state. The personal key never enters this view.

Outside authentication or commitment failures, every uncorrupted sign-share
query corresponds to a matching approval of the same $m$.
If $\mathsf{Bypass}$ occurs, fewer than $t-f$ uncorrupted signing oracles
contributed to $m$, yet $V$ accepted a valid signature. This violates the
assumed per-message threshold unforgeability.
A union bound over the cryptographic failure events defines
$\epsilon_{\rm crypt}(\lambda)$. Unique encoding and execution of signed fields
exclude substitution; audience, epoch, expiry, and atomic nonce checks exclude
cross-context use and replay under the stated system assumptions. Atomic
request consumption additionally excludes execution of multiple authorized
attempts for the same one-shot request.
\end{proof}

Reviewer-level auditability comes from authenticated approval logs, verified
share records, and execution receipts; the final BLS signature contains a
single group element. The corruption threshold is tight for key protection:
at $f\ge t$, an attacker can reconstruct $\sk$ and sign arbitrary messages.
Semantic errors follow a different threshold, determined by the approval
decisions analyzed next.

\subsection{Proposal and Reviewer Errors}

A final review attempt fixes the canonical action, evidence package, world
state, and prior history before the remaining reviewer randomness is sampled.
Let $h$ include the authenticated policy, supplied evidence, and all prior
search, feedback, and review interactions. Define
\[
 B=\{I_\pi(m,w)=0\},\qquad
 p=\Pr(B),\qquad
 q_i(m,w,h)=\Pr(X_i=1\mid B,m,w,h)\le q,
\]
where $X_i$ records a false approval and released share from an uncorrupted
domain. If no candidate is produced, $B$ is false.
The bound applies uniformly to the candidates and histories selected by the
adversary. The $f$ fully controlled domains may sign arbitrarily. For an
unrestricted adversarial proposer, $p=1$ gives the worst case.

Define the binomial upper tail
\[
 T_{N,k}(z)=\sum_{j=k}^{N}\binom Nj z^j(1-z)^{N-j},
 \qquad T_{N,k}=0\ (k>N),\quad T_{N,k}=1\ (k\le0).
\]

\begin{proof}[Proof of Theorem~\ref{thm:semantic}]
Let $N=n-f$, $k=t-f$, and $X=\sum_{i\notin C}X_i$.
Outside $\mathsf{Bypass}$, an unsafe acceptance requires $B$ and $X\ge k$,
even if all corrupt domains supply shares. Thus
\[
 \Pr(\mathsf{Unsafe})\le
 p\,\Pr(X\ge k\mid B)+\epsilon_{\rm crypt}.
\]
Conditional on any fixed eligible $(m,w,h)$, independent Bernoulli variables
with parameters $q_i\le q$ are stochastically dominated by
$\operatorname{Bin}(N,q)$: draw independent uniforms $U_i$ on $[0,1]$ and
realize $X_i=\mathbf 1[U_i\le q_i]\le\mathbf 1[U_i\le q]$. Taking the expectation over $(m,w,h)$
gives~\eqref{eq:unsafe-independent}.
Without independence, $\mathbb E[X\mid B,m,w,h]\le Nq$.
Markov's inequality and the probability bound of one give
$\Pr(X\ge k\mid B,m,w,h)\le\min(1,Nq/k)$; averaging proves
\eqref{eq:unsafe-dependent}.
\end{proof}

The binomial term is attained in the idealized case of equal conditional
error rates $q$, all corrupt domains signing unsafe candidates, and guaranteed
submission and acceptance of every quorum-qualified candidate, with
cryptographic failure excluded. Extra verifier checks or withholding can
reduce acceptance. A Poisson-binomial tail gives the corresponding expression
for heterogeneous independent error rates.

The dependence bound is tight given only the marginal rates.
When $Nq\le k$, with probability $Nq/k$ choose a uniformly random $k$-subset
of the uncorrupted reviewers to err; otherwise none err. Every marginal is
$q$, but the threshold fails with probability $Nq/k$. When $Nq\ge k$,
mix subset sizes $\lfloor Nq\rfloor$ and $\lceil Nq\rceil$ to achieve mean
$Nq$; every selected subset reaches $k$ and every marginal is again $q$.
For the simpler common-error example $X_1=\cdots=X_N=Z$,
$Z\sim\operatorname{Bernoulli}(q)$, threshold failure is $q$ for every
$1\le k\le N$. Computing the threshold tail more precisely requires a joint
error model, such as the Gaussian construction in Appendix~\ref{app:evaluation}.

\paragraph{Relation to prompt injection.}
Prompt injection is one way to induce the errors in this model. \textsf{KITA} is orthogonal to prompt-injection defenses: such defenses lower
reviewer error rates, while \textsf{KITA} limits what those errors can authorize.
Combining both can improve the security of agentic systems.
Multiple reviewers can reduce unsafe execution relative to a proposer with
unilateral signing authority when their joint errors and the threshold
support that reduction. Even if the quorum is misled, its approvals release
only action-bound signature shares. The delegated key remains protected
under Theorem~\ref{thm:integrity}'s assumptions with fewer than $t$ fully
corrupt domains.

\subsubsection{Conditional bounds and empirical error rates}
\label{app:measurement}
Write $x=(m,w,h)$ for an unsafe review instance. Conditional on $x$, randomness
comes from the review procedure, such as model sampling or randomized
retrieval. A deterministic procedure has error probability zero or one on a
fixed instance. The theorem assumes $\sup_x q_i(x)\le q$ over the stated
candidate/history space. The proposer rate $p$ is defined over the proposal
process and workload; an unrestricted attacker can choose only unsafe inputs.

An empirical review error rate averages over a specified distribution
$D$ of unsafe instances. Even if reviewers have equal error probability
$q(x)$ and are independent conditional on $x$, the committee rate under ideal
submission is $\mathbb E_{x\sim D}[T_{n,t}(q(x))]$ for $f=0$.
Substituting $\bar q=\mathbb E_D[q(x)]$ into $T_{n,t}$ generally changes this
quantity. For example, suppose all reviewers always err on 20\% of unsafe
instances and always reject the rest. Conditional outcomes are degenerate
and independent, each reviewer's mean error is $0.2$, and the $2$-of-$3$
committee error is $0.2$. Substitution gives $T_{3,2}(0.2)=0.104$.
The uniform bound for this population is $q=1$.

Repeated reviews of a fixed instance characterize within-instance randomness;
averaging over tasks also introduces variation in input difficulty. An
adaptive proposer can concentrate requests on difficult instances, changing
the distribution relative to an evaluation set. The uniform per-history
assumption on $q$ covers this selection in the theorem.

\subsection{Timely Completion}
\label{app:availability}

For an admitted valid fixed review package, let $Y_i=1$ mean that uncorrupted
domain $i$ delivers a valid approval share to the combiner by deadline $D$. Assume
$\Pr(Y_i=0\mid m,w,h,I_\pi=1)\le r$ uniformly over eligible valid candidates
and histories. This bound covers false rejection, abstention, and service or
channel failure before the deadline. Completion therefore depends on a
separate error parameter from the false-approval bound $q$.

\begin{samepage}
\begin{proposition}[Timely completion]
\label{prop:availability}
Assume that, after $t$ valid shares are available, combination, forwarding,
verification, and execution finish within $\Delta$. The request remains
pending and valid, with no conflicting execution, policy revocation, key
retirement, or expiry before $D+\Delta$. If the $Y_i$ are conditionally independent,
then, even with all $f$
corrupt domains withholding,
\begin{equation}
 \Pr(\mathsf{Complete}_{D+\Delta}\mid\hbox{admitted valid candidate})
 \ge T_{n-f,t}(1-r).
 \label{eq:availability}
\end{equation}
Without independence, for $t\le N=n-f$ a lower bound is
$\max\{0,1-Nr/(N-t+1)\}$; for $t>N$ the worst-case lower bound is zero.
\end{proposition}
\end{samepage}

\begin{proof}
Withholding leaves $N$ available uncorrupted domains; at least $t$ of them
must deliver shares by $D$. Independent success indicators with probabilities
at least $1-r$ dominate $\operatorname{Bin}(N,1-r)$. The downstream latency
bound then gives completion by $D+\Delta$.
Without independence, failure requires at least $N-t+1$ missing shares;
their expected number is at most $Nr$. Apply Markov's inequality to that count.
If $t>N$, corrupt withholding alone blocks the request.
\end{proof}

Let $a$ be the probability that an attempt produces and admits a valid
candidate to review. Under the timing and eligibility assumptions, total timely valid
completion is at least $aT_{n-f,t}(1-r)$.
In general $a+p\le1$, since an agent can also stall.
Reliable forwarding is necessary because a malicious $P$ can suppress the
request or completed signature, and an adversarial network can delay delivery
indefinitely. To tolerate $f$ corrupt domains for both integrity and
worst-case availability, choose $f<t\le n-f$, requiring $n\ge2f+1$.

For the special case $f=0$ with equal bounds $r=q$, no outages or stalls,
and conditionally independent errors, unanimity gives an unsafe-acceptance
upper bound of $pq^n+\epsilon_{\rm crypt}$ and a timely-valid-completion lower
bound of $(1-p)(1-q)^n$, under the timing and eligibility assumptions above
and guaranteed admission of valid candidates.
Equal bounds on false approval and false rejection need not be attained by
the review procedure.

\subsection{Adaptive Feedback and Multiple Attempts}
\label{app:retries}

Search and negotiation may contain many feedback rounds before $P$ selects
a candidate. They contribute to history $h$ and can influence the final
review errors, but produce no signature shares. The attempt budget $M$ counts
registered final authorization reviews. Any changed action or evidence
submitted for final approval, or request for freshly sampled final decisions,
uses a new attempt ID and nonce. The request registry in
Appendix~\ref{app:object} binds each admitted attempt to one digest and enforces
the budget before uncorrupted domains review it. Uncorrupted harnesses durably
cache approve, reject, and revise decisions for the same message digest and
do not re-sample them. This prevents a proposer from pooling partial approvals
from repeated trials on the same signed bytes. Fresh decisions require a
different message, so shares from different attempts cannot be combined.
Caching fixes the sampling unit for the analysis: one decision per reviewer
on each signed message. The per-attempt $q$ assumption holds conditional on
all earlier outcomes, including failed attempts.
For at most $M$ attempts with a fixed corruption set and uniform conditional
bounds $p_j,q_j$ at attempt $j$, the tower property and union bound give
\[
 \Pr\!\left(\bigcup_{j=1}^{M}F_j\right)
 \le \min\!\left\{1,\,
 \sum_{j=1}^{M}p_jT_{n-f,t-f}(q_j)
 +\epsilon_{\rm crypt}^{(M)}(\lambda)\right\}.
\]
The dependence-robust result substitutes $\min(1,(n-f)q_j/(t-f))$ for each
tail. The union bound applies to dependent attempts. Cryptographic failure is
bounded over the entire execution; the horizon must be polynomial in $\lambda$.
The expression $1-(1-s)^M$ is exact only with independent attempts of equal
failure probability $s$. Rate limits, expiration, and request-level attempt
budgets are therefore relevant even though nonces already prevent replay.

\FloatBarrier

\section{Threshold BLS and Key Lifecycle}
\label{app:lifecycle}

\subsection{Threshold BLS Instantiation}
\label{app:bls}

\paragraph{Base signature.}
BLS uses groups of prime order $\ell$,
$\mathbb{G}_1$, $\mathbb{G}_2$, and
$\mathbb{G}_T$ with a bilinear pairing
$e:\mathbb{G}_1\times\mathbb{G}_2\rightarrow\mathbb{G}_T$~\citep{bls2001}.
Let $g_2$ generate $\mathbb{G}_2$ and let
$H_{\mathrm{DST}}$ hash canonical bytes to $\mathbb{G}_1$ under a
protocol-specific domain-separation tag. For secret $x\in\Zq$, the public key,
signature, and verification equation are
\begin{equation}
  \pk=g_2^x,\qquad h=H_{\mathrm{DST}}(m),\qquad
  \sigma=h^x,\qquad e(\sigma,g_2)=e(h,\pk).
\end{equation}
The signature is deterministic and contains one group element. Our
minimal-signature-size BLS12-381 profile encodes $\sigma\in\mathbb{G}_1$ in 48
bytes and $\pk\in\mathbb{G}_2$ in 96 bytes; the opposite placement reverses
these sizes~\citep{blsdraft2026}. The prototype computes $H_{\mathrm{DST}}$
with py\_ecc's \texttt{hash\_to\_G1}, which implements the RFC~9380 suite
\texttt{BLS12381G1\_XMD:SHA-256\_SSWU\_RO\_}, under the application tag
\texttt{AGENTAUTH\_BLS12381G1\_XMD:SHA-256\_SSWU\_RO\_}~\citep{rfc9380,pyecc}.

\paragraph{Threshold construction.}
Threshold BLS applies Shamir sharing~\citep{shamir1979} in the exponent~\citep{boldyreva2003}.
Trusted user provisioning or verifiable DKG produces shares of a random
degree-at-most-$(t-1)$ polynomial $g$ with $g(0)=x$.
Only the dealer profile temporarily constructs $x$ in one trusted location.
Signer $S_i$ stores $x_i=g(i)$ and publishes an
authenticated verification share $\pk_i=g_2^{x_i}$. After its paired reviewer
approves the exact $m$, it returns $\sigma_i=h^{x_i}$. The combiner first checks
$e(\sigma_i,g_2)=e(h,\pk_i)$, then for any set $S$ of $t$ valid indices computes
\begin{equation}
  \lambda_i=\prod_{j\in S,\,j\ne i}\frac{-j}{i-j}\pmod\ell,
  \qquad
  \sigma=\prod_{i\in S}\sigma_i^{\lambda_i}=h^x.
\end{equation}
Thus, each signature share can be produced independently in the online phase,
shares on different messages cannot combine, and the verifier receives the same
one-signature interface as ordinary BLS.

\paragraph{Online signing.}
BLS gives each isolated signer a one-request, one-response online path without a
distributed nonce round, while the final authorization remains one group
element as the committee grows. The deployment profile pairs BLS with secure setup,
authenticated public shares, canonical encoding, domain separation, standard
hash-to-curve, and strict point validation~\citep{rfc9380,blsdraft2026}.

\subsection{Key Setup and Lifecycle}

\paragraph{Setup.}
In the user-provisioned profile, $U$ generates a fresh delegated key with
cryptographic randomness, splits it into Shamir shares, and sends each share
to its designated protected signer over a confidential authenticated channel.
$U$ authenticates the committee, threshold, public verification shares, group
key, and policy/key epoch to the relevant parties. The complete delegated key,
polynomial, and recoverable setup copies are erased after provisioning.
The personal key remains in the user-controlled wallet and is not shared.
Alternatively, $U$ authorizes verifiable DKG among the isolated services
$S_1,\ldots,S_n$~\citep{gennaro2007dkg}. Each signer obtains only its share, so no complete delegated
key is assembled. 
\paragraph{Epochs and recovery.}
Shares occupy separate failure domains. A key epoch identifies one independently
generated delegated key and its sharing configuration. Rotation or committee
reconfiguration provisions a fresh key, retires the old key at $V$, and erases
old shares. Policy updates change the authenticated policy version; keeping
the same key also keeps its corruption budget, including all previously
exposed shares. Relabeling an epoch does not reset that budget. Request and
attempt state survives either update.
Execution requires $t$ available signers, so loss of quorum suspends
authorization until recovery or reconfiguration. Proactive refresh with secure
erasure~\citep{ostrovsky1991proactive,herzberg1995proactive} would require a separate analysis. The present guarantee uses a fixed
corruption set over the lifetime of each key epoch.

\paragraph{Signature compatibility.}
BLS provides a noninteractive online share-generation step and a compact final
signature. An executor with an ECDSA interface instead needs a threshold-ECDSA
implementation~\citep{lindell2017,gg18,cmp2020}, including its nonce and coordination protocol. Reconstructing
the group secret in a conventional signer would restore a complete-key
custodian. The deployment choice follows the executor's signature interface.

\FloatBarrier
\section{Application to Agent Payments}
\label{app:ap2}

\subsection{Illustrative Purchase Review}
\label{app:review-task}
Consider a user buying a new laptop for a named CAD application with a
\$1,500 total budget. The authenticated request names the application version,
requires vendor-supported operation, and excludes refurbished units.
Table~\ref{tab:review-task} separates the checks needed for this purchase.
Model-based review interprets product descriptions and supporting documents
when the relevant facts lack normalized fields. Deterministic checks enforce
constraints on the resulting structured fields.

\begin{table}[ht]
\centering\small
\setlength{\tabcolsep}{4pt}
\begin{tabular}{L{0.19\linewidth}L{0.35\linewidth}L{0.38\linewidth}}
\toprule
Check & Purchase requirement & Review operation \\
\midrule
Deterministic & Total including fees $\le\$1,500$; permitted payee and currency & Compare canonical fields with authenticated policy. \\
Semantic & New unit; supported configuration for the named application & Interpret condition qualifiers and match exact SKU/OS to vendor support documentation. \\
Evidence & Order identity, provenance, and freshness & Verify checkout signatures; retrieve approved sources; resolve conflicts or request more evidence. \\
\bottomrule
\end{tabular}
\caption{Illustrative purchase requirements and corresponding review operations.}
\label{tab:review-task}
\end{table}

A \$1,299 listing may pass the amount and merchant checks yet describe a
``like-new, renewed'' unit. A reviewer can flag this during discussion; $P$
then searches for a new unit, supplies another listing, and asks reviewers
to resolve remaining compatibility questions. Once $P$ selects a candidate,
it freezes the order and evidence and requests final approval. If the issue
is discovered only during final review, the reviewer returns revise without
a signature share; the replacement enters a new authorization attempt.
Approval binds the SKU, condition, quantity,
total, policy reference, and evidence digest. This example distinguishes two
failures: misreading ``renewed'' is a semantic error; substituting a different
SKU after approval violates authorization integrity.

\subsection{Mandate and Execution Bindings}

AP2's autonomous mode binds an Agent Key to a user-authorized open mandate.
A \textsf{KITA} integration would bind the committee group key in its place, with the
retailer or payment service acting as the trusted executor $V$. This proposed
mapping adds payment-specific bindings to the protocol in
Appendix~\ref{app:object}.

The open mandate specifies the authorized key and purchase limits. A
merchant-signed checkout fixes the SKU, condition, quantity, payee, currency,
total including fees, and order identifier. Reviewers validate its provenance
and assess the supporting evidence against the authenticated requirements.
The resulting closed mandate $m$ includes these order fields, the checkout
digest, and a reference to the open mandate. A signature therefore covers
both the reviewed purchase and its user-authorized scope.

Before execution, the payment tool verifies the mandate chain, merchant and
committee signatures, audience, key epoch, policy version, expiry, registered
attempt, and pending request ID. Atomic or idempotent execution consumes the
request as well as the nonce and binds the verified order to the payment. Normalization
that changes signed fields requires a new message and review round;
settlement and receipts remain responsibilities of the payment stack.

The mapping also requires signature compatibility. The referenced AP2 v0.2
examples use JOSE ES256 (P-256) signatures~\citep{rfc7515,rfc7518}. Threshold
ECDSA~\citep{gg18,cmp2020} can preserve that signature interface; BLS requires an extended profile specifying the
algorithm, key encoding, and canonical signing input.

\section{Validation Methods and Supplementary Results}
\label{app:evaluation}

\subsection{Prototype and Model Integration}

The implementation combines a typed reviewer interface, authenticated approval
gateways, BLS signer services, a share combiner, and an executor within one
Python process. Each reviewer receives the same canonical message, policy,
and evidence, and returns an outcome with the message digest, reason codes,
and rationale. Gateways authenticate matching approvals; signers verify the
authentication and digest before releasing shares. The executor verifies the
group signature, expiry, and nonce. Experiments use BLS12-381 signing and verification. The prototype runs one
final review per candidate; the negotiation phase, durable decision cache,
attempt registry, cross-attempt request consumption, and key-rotation lifecycle
are specified deployment components outside these tests.

Reviewer inputs contain public review data and exclude signing shares and
approval-channel credentials. Scripted decisions exercise the authorization
checks independently of model behavior. A single-request integration test
used three separate sessions of the \texttt{gpt-5.6-sol} model alias, invoked
in September 2026 through the Codex CLI's non-interactive structured-output
mode with medium reasoning effort, identical inputs, and default sampling settings; temperature and seed controls
were unavailable. The benign request specified a USD 1,299 laptop under a
USD 1,500 limit, with permitted merchant and currency and fixture-based
merchant-signature validation. All reviewers approved. The pipeline executed
the approved action once and rejected modification and replay.
This live-model run is separate from the six scripted system tests below.

\subsection{Scripted System Tests}
\label{app:system-tests}

The system tests replace each LLM with a scripted reviewer at the same typed
interface, so they exercise the authorization path rather than model accuracy.
Each test builds a fresh $2$-of-$3$ committee with dealer-generated shares.
Table~\ref{tab:system-tests} lists the six tests; all pass in the released
artifact.

\begin{table}[ht]
\centering\small
\setlength{\tabcolsep}{4pt}
\begin{tabular}{rL{0.24\linewidth}L{0.30\linewidth}L{0.33\linewidth}}
\toprule
\# & Test & Setup & Required outcome \\
\midrule
1 & Quorum authorizes & Two approvals, one rejection & Combined signature verifies; $V$ executes. \\
2 & Insufficient quorum & One approval, two rejections & One valid share; no signature is formed. \\
3 & Digest mismatch & One reviewer approves a different digest & Gateway rejects that approval; quorum is not reached. \\
4 & Post-review substitution & Amount changed from 1,299 to 1,499 after signing & $V$ rejects the signature on the modified action. \\
5 & Replay & Accepted action resubmitted & Spent nonce blocks the second execution. \\
6 & Key-free identical input & Evidence containing injected instructions & All reviewers receive identical bytes with no key or channel-secret fields. \\
\bottomrule
\end{tabular}
\caption{Scripted system tests of the in-process prototype.}
\label{tab:system-tests}
\end{table}

\subsection{Fault-Injection Validation Across Baselines}
\label{app:baseline-tests}

We instantiate the four designs in Table~\ref{tab:baselines} using BLS for
execution signatures and HMAC for reviewer authentication. The quorum is
$2$-of-$3$ where applicable. Every executor is trusted and enforces expiry
and one-time nonces. Scripted decisions represent correct or mistaken reviews
of actions whose payment fields satisfy the fixture's structural constraints;
the unsafe label represents an additional semantic requirement.

Each proposer uses the credentials available to its design. A breach of one
custody domain exposes the agent-held or central signer's complete key, one
authentication credential in direct approval verification, or one key share
and approval credential in \textsf{KITA}; other reviewers reject. The intervention
measures the effect of each credential placement, with exposure assigned
directly rather than obtained through an exploit. Substitution changes the
amount from 1,299 to 1,399 after approval, and replay resubmits an accepted
message. Each scenario begins with fresh verifier state.
Table~\ref{tab:baseline-results} reports the resulting execution decisions.

\begin{table}[ht]
\centering\small
\setlength{\tabcolsep}{5pt}
\begin{tabular}{lcccc}
\toprule
Scenario & Agent key & Central signer & Direct approvals & \textsf{KITA} \\
\midrule
Valid, two approvals & Accept & Accept & Accept & Accept \\
Unsafe, zero approvals & Accept & Block & Block & Block \\
Unsafe, one false approval & Accept & Block & Block & Block \\
Unsafe, two false approvals & Accept & Accept & Accept & Accept \\
One custody domain breached & Accept & Accept & Block & Block \\
Post-review substitution & Accept & Block & Block & Block \\
Replay after acceptance & Block & Block & Block & Block \\
\bottomrule
\end{tabular}
\caption{Execution outcomes under controlled review decisions and credential
exposure. The custody row uses the scheme-specific compromise defined above.
Two mistaken approvals authorize an unsafe action in all three quorum designs.}
\label{tab:baseline-results}
\end{table}

\subsection{Committee Sensitivity Model}
\label{app:simulation}

\paragraph{Analytical example.}
The bounds quantify the cost of a stricter quorum under assumed error rates.
For $n=3$, $f=0$, $p=0.1$, $q=0.2$, and $r=0.1$, changing from $t=2$ to
$t=3$ reduces the independent-error unsafe bound from $0.0104$ to $0.0008$,
while the lower bound on conditional valid completion falls from $0.972$ to $0.729$.
If one domain withholds, unanimous completion becomes impossible.
The marginal-only unsafe bounds are $0.03$ and $0.02$, respectively:
the guarantee from marginal error bounds alone is substantially weaker.

\paragraph{Correlated review model.}
For each request class, reviewer $i$ has latent score
\begin{equation}
  L_i=\sqrt{\rho}\,Z+\sqrt{1-\rho}\,\epsilon_i,
  \qquad \mathsf{approve}_i=\mathbf{1}[L_i>\tau],
\end{equation}
where $Z$ and the $\epsilon_i$ are independent standard Gaussian variables. The
threshold $\tau=\Phi^{-1}(1-\alpha)$ gives marginal approval probability
$\alpha$.
We set $\alpha=0.90$ for legitimate requests and $\alpha=0.20$
for out-of-policy requests, with $f=0$ and no outages; Figure~\ref{fig:committee}
writes these marginal rates as $1-\bar r$ and $\bar q$. They average over $Z$
and are not the uniform conditional bounds of Theorem~\ref{thm:semantic} and
Proposition~\ref{prop:availability}. For $\rho>0$, conditional on a large $Z$
every reviewer approves an out-of-policy request with probability near one, so
the uniform bound for this population is $q=1$
(Appendix~\ref{app:measurement}). 
The parameter $\rho$ controls latent-score correlation. We separately estimate
the induced correlation between binary votes from the simulated approval count.

For committee size $n$ and authorization threshold $t$, the error rates are
\begin{equation}
 \mathsf{FA}=\Pr\!\left[\textstyle\sum_i \mathsf{approve}_i\ge t
   \mid\text{out of policy}\right],\quad
 \mathsf{FR}=\Pr\!\left[\textstyle\sum_i \mathsf{approve}_i<t
   \mid\text{legitimate}\right].
\end{equation}
Each estimate uses $N_{\rm MC}=400{,}000$ trials with NumPy generator seed
20260720. Monte Carlo uncertainty under this distribution is summarized by
the normal binomial interval
$\hat\theta\pm1.96\sqrt{\hat\theta(1-\hat\theta)/N_{\rm MC}}$.
For the $n=7$, $\rho=0.2$ threshold sweep, the largest 95\% interval
half-width is below $0.0016$. The induced binary correlations are $0.106$
for out-of-policy requests and $0.080$ for legitimate requests.
The reproducibility artifact retains the full sweep and intervals.

Under ideal submission and execution, committee FA translates to per-attempt
unsafe acceptance $p\,\mathsf{FA}$. With valid-candidate rate $a$ and reliable
forwarding, valid completion is $a(1-\mathsf{FR})$ in this no-outage model.

\paragraph{Threshold and dependence effects.}
At latent correlation $\rho=0.2$, a $4$-of-$7$ threshold yields FA $0.083$
and FR $0.017$, compared with $0.20$ and $0.10$ for one reviewer. Unanimity
reduces FA to $0.0013$ but raises FR to $0.45$.
Figure~\ref{fig:committee} shows the trade-off across thresholds and correlations.
At $t=4$, FA rises from $0.033$ under independence toward $0.20$ as
$\rho\rightarrow1$. For the modeled rates at $\rho=0.2$, an intermediate
threshold improves both errors over a single reviewer; unanimity sacrifices
valid completion to reduce false approval further. Selecting a threshold
therefore requires joint-error estimates and a tolerance for rejected valid
requests.

\subsection{Cryptographic Microbenchmarks}
\label{app:benchmarks}

We measure share generation, share verification, combination, and final
verification using py\_ecc 8.0.0~\citep{pyecc} on Python 3.9.6 and macOS 15.7.4 on Apple
Silicon. Dealer-generated Shamir shares use a deterministic PRNG with seed 7.
The configurations span committee sizes from 3 to 32 with thresholds
between $0.6n$ and $0.72n$; they were chosen to cover committee sizes rather
than to follow one threshold rule. Per-share signing, per-share verification,
and final verification do not depend on $(n,t)$, while combination grows with
$t$, so a $4$-of-$7$ committee as in Figure~\ref{fig:committee} combines in time
between the $t=3$ and $t=5$ rows. The combiner verifies each of its $t$ shares,
so its total verification work also grows linearly in $t$.
Each $(n,t)$ configuration signs one message, verifies admitted shares,
combines $t$ shares, and verifies the final signature. Share generation and
combination use 25 repetitions; pairing-based verification uses five.
Table~\ref{tab:blsbench} reports arithmetic means for this pure-Python
implementation.

\begin{table}[ht]
\centering
\small
\setlength{\tabcolsep}{5pt}
\begin{tabular}{rrrrrr}
\toprule
$n$ & $t$ & share sign (ms) & share verify (ms) & combine (ms) & final verify (ms) \\
\midrule
3  & 2  & 4.907 & 558.237 & 5.045  & 586.684 \\
5  & 3  & 5.708 & 565.976 & 5.536  & 561.076 \\
7  & 5  & 5.058 & 567.418 & 11.367 & 555.983 \\
10 & 7  & 5.054 & 558.328 & 16.724 & 554.668 \\
16 & 11 & 5.117 & 560.180 & 28.711 & 565.716 \\
24 & 17 & 4.936 & 556.257 & 47.746 & 567.630 \\
32 & 22 & 5.042 & 558.746 & 62.039 & 555.971 \\
\bottomrule
\end{tabular}
\caption{Threshold-BLS operation times in milliseconds. All configurations use
a 48-byte signature, a 96-byte public key, and noninteractive signature-share
generation.}
\label{tab:blsbench}
\end{table}

\end{document}